\documentclass[11pt]{article}

\usepackage{amssymb}
\usepackage{amsmath}
\usepackage{bm}
\usepackage{graphicx}
\usepackage{amsthm}
\usepackage{xcolor}
\usepackage{bbm} 

\def\x{{\bm{x}}}
\def\y{{\bm{y}}}

\def\W{{\bm{W}}}
\def\b{{\bm{b}}}

\def\R{{\mathbb{R}}}
\def\dist{\mathop{\mathrm{dist}}}
\def\btheta{{\mbox{\boldmath$\theta$}}}

\def\dive{\mathop{\mathrm{div}}}

\title{Signed Distance Function Computation from an Implicit Surface}
\author{Pierre-Alain Fayolle\\
Computer Graphics Laboratory, University of Aizu,\\
Aizu-Wakamatsu, Japan,\\
fayolle@u-aizu.ac.jp}
\date{}

\begin{document}

\maketitle

\begin{abstract}
We describe in this short note a technique to convert an implicit surface into a Signed Distance Function (SDF) while exactly preserving the zero level-set of the implicit. 
The proposed approach relies on embedding the input implicit in the final layer of a 
neural network, which is trained to minimize a loss function characterizing the SDF. 
\end{abstract}

\section{Introduction}
We consider in this note the problem of computing the Signed 
Distance Function (SDF) to an implicit surface. 
Let a surface $S$ be defined as the zero level-set of a 
function $f$, $S = \{\x\in\R^3: f(\x)=0\}$. We propose a 
method for computing the signed distance to $S$: 
\begin{equation}\label{dist}
d(\x) = \pm\dist(\x, S) = \pm \min_{\y \in S} |\x - \y|, 
\end{equation}
while preserving exactly the zero level-set of $f$. 
W.l.o.g we assume that $d>0$ inside the domain bounded by $S$
and that $d<0$ outside. 

In the literature on level-set methods, techniques for computing 
the SDF $d$ from $f$ are called re-initialization 
methods \cite{sussman1994level,osher2006level}. 
One considers the problem: 
\begin{equation}\label{reinit}
\partial g(\x,t) / \partial t + 
(1 - |\nabla_{\x} g(\x,t)|) = 0 \qquad g(\x,0) = f(\x)
\end{equation}
and solves it to steady-state. 
Typically, (\ref{reinit}) is solved by discretizing 
the time derivative with explicit or implicit Euler, 
and the spatial derivative by finite difference
\cite{sussman1994level}. 
Another popular approach for computing (\ref{dist}) 
is the Fast Marching Method \cite{sethian1996fast}. 
\\
In each case, the approach relies on computing a 
regular grid for the domain, and sampling functions and 
derivatives on this grid. Thus the zero level-set 
of $g$ does not correspond exactly to the zero 
level-set of $f$. 
\\
On the other hand, the method described here preserves 
exactly the zero level-set of $f$. It does so by computing 
the signed distance $d(\x;\btheta)$ as $f(\x)g(\x;\btheta)$, 
where $g(.)$ is a parametric function whose parameters 
$\btheta$ are fitted such that $d$ corresponds to the distance function.

\section{Related work}
\paragraph{Distance to an implicit surface}
The simplest method for turning $f$ into a SDF consists in 
meshing the zero level-set $S$, for example with the Marching Cubes 
algorithm \cite{lorensen1987marching}, 
and then to compute the distance to the polygonal mesh. 
Typically accelerating data-structures, 
such as a Kd-Tree or a Bounding Volume Hierarchy (BVH), are used 
for practical computations. 

Re-initialization (or re-distancing) a scalar field $f$ sampled 
on a grid is a common operation in level-set methods
\cite{sussman1994level}. See, as well, \cite{osher2006level}. 
It relies on solving (\ref{reinit}) by finite differences 
for the spatial derivatives and explicit Euler for time-stepping. 
Several variants were proposed over the years that use higher-order 
schemes for approximating the spatial derivatives or for 
time-stepping. 
A related approach is the Fast Marching Method of Sethian 
\cite{sethian1996fast}. 
\\
All these approaches rely on sampling the function and its 
derivatives on a grid (generally a regular grid, but methods 
working on hierarchical grids exist as well). 

\paragraph{Deep neural networks for computational science}
While deep neural networks were originally designed for tackling 
problems of regression or classification, their use has now 
been extended to deal with different sorts of problems in 
computational sciences and applications. 
\\
Deep neural networks can be trained for computing the SDF to 
a given point-cloud or triangle mesh 
\cite{atzmon2019controlling,atzmon2020sal,atzmon2020sal++,chibane2020neural,gropp2020implicit,davies2021effectiveness,sitzmann2020implicit,sitzmann2020metasdf,takikawa2021nglod}. 
All these methods differ in how they model the loss function 
or the network architecture. 
For example, \cite{sitzmann2020implicit} proposes to use 
$\sin()$ activation functions. 
\\
In this work, instead of starting from a point-cloud or 
a polygonal mesh, we assume that the input is a function $f$ 
whose zero-level set describes the surface of interest. 
Information about the surface is embedded in the neural network 
by multiplying the final layer by the function $f$ itself (or 
by a smoothed version of $\mathrm{sign}(f)$). 

Related to our approach are recent techniques for solving 
Partial Differential Equations (PDE), including high-dimensional 
and stochastic PDE, \cite{weinan2018deep,blechschmidt2021three} 
with deep neural networks.

\section{Approach}
Given a function $f$ whose zero level-set defines a surface $S$, 
our goal is to compute $d$, the signed distance to 
$S$ (\ref{dist}). 
We look for the distance function $d$ under the form 
\begin{equation}\label{ansatz}
d(\x;\btheta)=f(\x) g(\x;\btheta)
\end{equation}
where $g$ is a parametric function with parameters $\btheta$. 
We use a fully connected feedforward neural network
for $g$. For fitting $\btheta$, the parameters of $g$, we express 
$d$ as the solution to a variational problem (See Sections \ref{sec:variational} 
and \ref{sec:loss}). 
\\
An alternative to (\ref{ansatz}) is to use 
\begin{equation}\label{ansatz2}
d(\x;\btheta) = \mathrm{sign}(f(\x)) g(\x;\btheta), 
\end{equation}
where $\mathrm{sign}(x)$ is a smoothed version of 
the sign function. 
For example, one can use $\mathrm{sign}(x) \equiv \tanh(\alpha x)$, 
where $\alpha$ is a user-specified parameter.

\subsection{Feedforward neural network}
We use as an ansatz for $d(\x)$, $f(\x)g(\x;\btheta)$, where $x_L=g(\x;\btheta)$ 
is a deep, fully connected neural network defined by 
\begin{align*}
    x_L &= W_L \, \sigma(\x_{L-1}) + b_L\\
    \ldots \\
    \x_2 &= W_2 \, \sigma(\x_1) + b_2\\
    \x_1 &= W_1 \, \x + b_1.
\end{align*}
$L$ corresponds to the number of layers in the deep 
neural network, $W_i$ and $b_i$ correspond to the parameters $\btheta$
of the neural network, and $\sigma()$ is a non-linear 
activation function. 

\subsection{Variational problems}
\label{sec:variational}
The distance function $d$ (\ref{dist}) is the viscosity solution 
to the eikonal equation \cite{crandall1983viscosity}
\begin{equation}\label{eik}
|\nabla_{\x} d| = 1
\end{equation}
with boundary condition $d(\x)=0$ for $\x \in S$. 
See, e.g. \cite{belyaev2015variational,belyaev2019admm} for 
recent numerical techniques to solve (\ref{eik}). 

An alternative approach consists in considering the following $p$-Poisson problem
\begin{equation}\label{pPoisson}
\Delta_p u = -1, 
\end{equation}
subject to $u=0$ on $S$, and, where 
$\Delta_p u \equiv \dive\left( |\nabla_{\x} u|^{p-2} |\nabla_{\x} u| \right)$
is the $p$-Laplacian. 
\\
Let $u_p$ be the solution to (\ref{pPoisson}), 
$u_p(\x) \rightarrow \dist(\x,S)$ as 
$p \rightarrow \infty$ and thus can be used to approximate 
the signed distance function to $S$ 
\cite{belyaev2015variational,fayolle2018p}. 
\\
For a fixed value of $p$, the solution $u_p$ to (\ref{pPoisson}) only 
delivers an approximation of the distance function to $S$. 
The approximation quality can be improved by a simple normalization 
scheme as described in \cite{belyaev2015variational}. 

\subsection{Loss functions}
\label{sec:loss}
Let $g(\x; \btheta=(\W,\b))$ be a deep, 
fully connected neural network with parameters $\btheta=(\W,\b)$. 
We use the ansatz $d(\x;\btheta)=f(\x)g(\x; \btheta)$ 
(or $d(\x;\btheta)=\mathrm{sign}(f(\x)) g(\x; \btheta)$)
for the distance function, where the parameters $\btheta$ need to be fitted. 
\\
Given that the distance function is the viscosity solution 
to the eikonal equation (\ref{eik}), we consider the following loss 
function (to be minimized)
\begin{equation}\label{loss_eik}
L(\btheta) = \mathbb{E}_{\x \sim D} \left(\left|\nabla_{\x} d(\x;\btheta)\right|-1\right)^2, 
\end{equation}
where $\mathbb{E}$ is the expected value, 
and $D$ is the uniform distribution over the given computational 
domain. 
\\
Similarly, if one wants to solve the $p$-Poisson problem (\ref{pPoisson}) instead, 
the correspond loss function is given by
\begin{equation}\label{loss_pPoisson}
L(\btheta) = \mathbb{E}_{\x \sim D} \left(\Delta_p d(\x;\btheta) + 1\right)^2.
\end{equation}
\\
It is possible to add other constraints to the loss (\ref{loss_eik}) or 
the loss (\ref{loss_pPoisson}). For example, if necessary, 
one could prevent eventual extra zero level-sets 
away from the surface by adding a term such as
\begin{equation}\label{add_constraint}
\mathbbm{1}_{\{\x:f(\x)\neq 0\}}\exp\left(-\gamma \, |d(\x;\btheta)|\right), 
\end{equation}
where $\mathbbm{1}_X$ is the indicator function for the set $X$, 
and $\gamma$ is a large value (for example $\gamma = 10^2$). 

Stochastic Gradient Descent \cite{robbins1951stochastic} 
or a variant, such as e.g. ADAM \cite{kingma2014adam}, is 
used for minimizing the loss function (\ref{loss_eik}) or 
the loss function (\ref{loss_pPoisson}), 
with eventual additional constraints such as (\ref{add_constraint}). 
\\
The corresponding minimizer $d(\x; \btheta=\widetilde{\btheta})$, 
with fitted parameters $\widetilde{\btheta}$ provides 
the SDF to $S$ (or an approximation in the case of 
(\ref{loss_pPoisson})), while preserving exactly the zero level-set of $f$. 

\subsection{Derivatives computation}
We rely on automatic differentiation to compute the 
spatial derivatives $\nabla_{\x}$, $\dive$, or $\Delta_p$ exactly. 
Note that the derivatives are taken here w.r.t. the 
input $\x$ of the function. 
\\
For training the deep neural network, we also 
need to compute the derivatives of the loss functions 
(\ref{loss_eik}) or (\ref{loss_pPoisson}) w.r.t. the parameters 
$\btheta=(\W,\b)$. This is also performed by automatic differentiation 
as usually done when training deep neural networks. 

\section{Numerical results}
\subsection{Setup}
We use Torch \cite{paszke_19} for specifying the deep 
neural network, computing the derivatives (w.r.t. the network's 
parameters and the coordinates), and minimizing the loss 
function. 

For the network $g$ we use 8 fully connected linear layers 
with width 512. We use a skip connection from the input to 
the middle layer. The architecture is similar to the one 
described in \cite{park2019deepsdf} and in \cite{gropp2020implicit}.
\\
The softplus activation function is used for the non-linear activation 
functions 
$$
\sigma(x) = \frac{1}{\beta}\ln(1+\exp(\beta x)).
$$
We used $\beta=100$ in the experiments. 
Using the softplus activation function behaved better in our experiments 
than the Rectified Linear Unit (RELU), which is typically used 
with deep neural networks. 
Note that it is also possible to consider alternatives such as 
the SIREN layer \cite{sitzmann2020implicit} for the activation function. 
We did not experiment with this alternative. 
\\
In the final layer, we multiply the output of the neural network 
$g(\x;\btheta)$ by $\tanh(\alpha f(\x))$, 
the smoothed sign function of $f(\x)$. We used $\alpha=0.1$. 
\\
We minimize the loss function (\ref{loss_eik}) 
(or (\ref{loss_pPoisson})) by ADAM \cite{kingma2014adam}, 
using a learning rate of $0.0001$. 
The parameters of the neural network are initialized with 
the geometric initialization of \cite{atzmon2020sal}.
\\
For the computation of the loss functions (\ref{loss_eik}) 
or (\ref{loss_pPoisson}), we use the uniform distribution 
over the computational domain for the distribution $D$. 

\subsection{Results}
\subsubsection{1D example}
We start our numerical experiments with a simple example in 1D: 
We consider the domain $[0,1]$ with boundary the points $\{0,1\}$. 
We build an implicit for this domain as
$$
f(x) = x + (1-x) + \sqrt{x^2 + (1-x)^2}, 
$$
i.e. the intersection of the half-spaces $x$ and $(1-x)$ 
with the intersection implemented by an R-function 
\cite{shapiro2007semi,pasko1995function}. 
\\
Note that in this case, the exact SDF to the boundary is known and
is given, of course, by 
$$
\min(x, 1-x).
$$ 
\begin{figure}[!h!tbp!]
\centering
\includegraphics[width=0.45\textwidth]{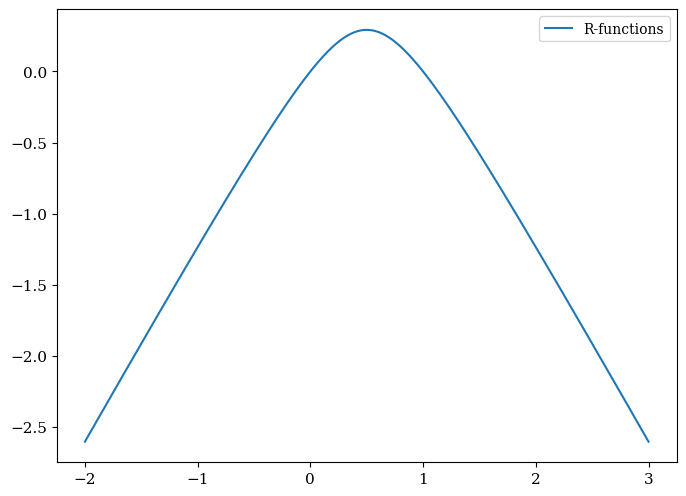}
\includegraphics[width=0.45\textwidth]{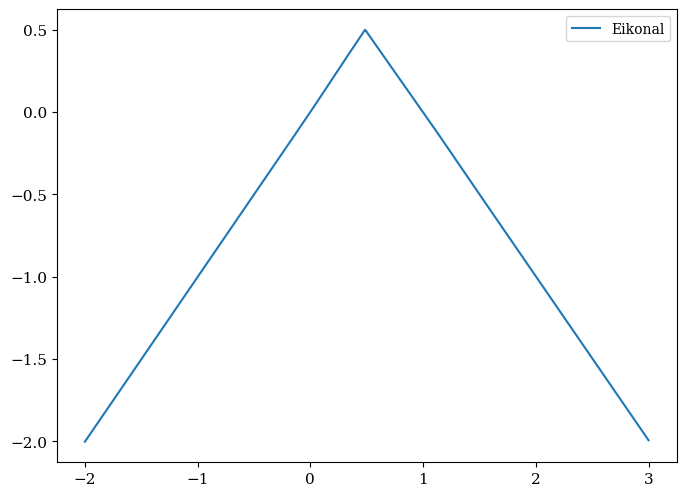}
\includegraphics[width=0.45\textwidth]{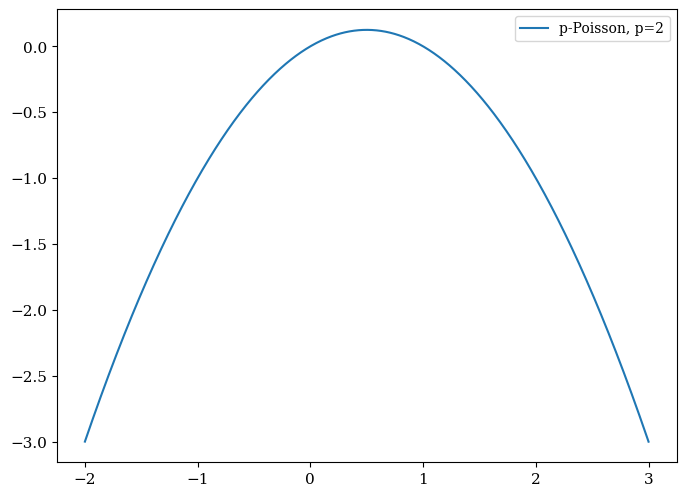}
\includegraphics[width=0.45\textwidth]{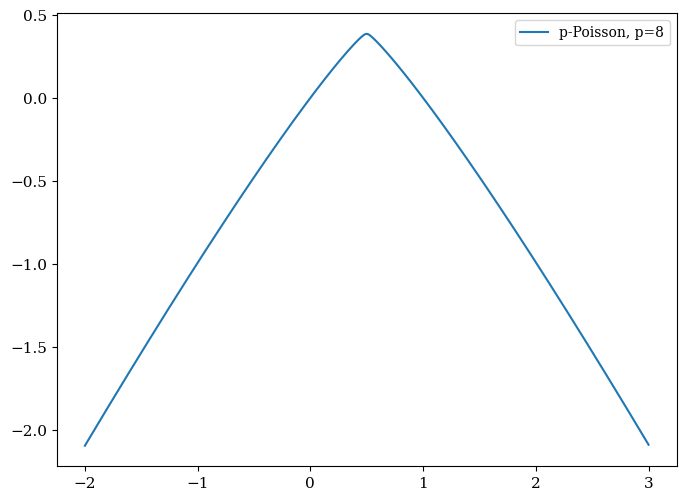}
\caption{Results in 1D.
Top row: 
Plot of $f(x)=x+(1-x)+\sqrt{x^2+(1-x)^2}$ (left image); 
Minimizing (\ref{loss_eik}), right image. 
And minimizing (\ref{loss_pPoisson}) with $p=2$, $p=8$ (left and right, 
respectively) in the bottom row. 
Note that the zero level-set of $f$ (the points $\{0,1\}$) is exactly preserved.}
\label{fig:1d_plot}
\end{figure}
Figure \ref{fig:1d_plot} shows a plot of the function $f$ (top-left image), 
the distance computed by solving the eikonal equation (\ref{eik}) in the top-right image,
and the approximate distance obtained by solving the $p$-Poisson problem (\ref{pPoisson})
with $p=2, 8$, respectively (bottom left and right images). 
\\
Solving (\ref{eik}) or (\ref{pPoisson}) is done by minimizing the corresponding 
loss function (\ref{loss_eik}) or (\ref{loss_pPoisson}), respectively. 
We used $15000$ iterations in both cases. 
\begin{figure}[!h!tbp!]
\centering
\includegraphics[width=0.31\textwidth]{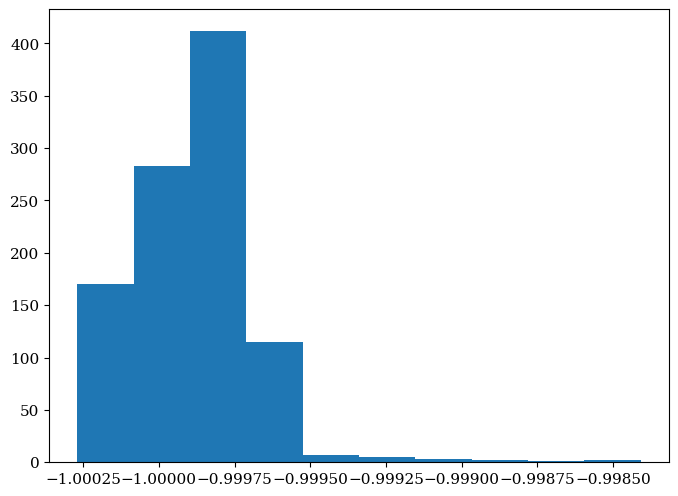}
\includegraphics[width=0.31\textwidth]{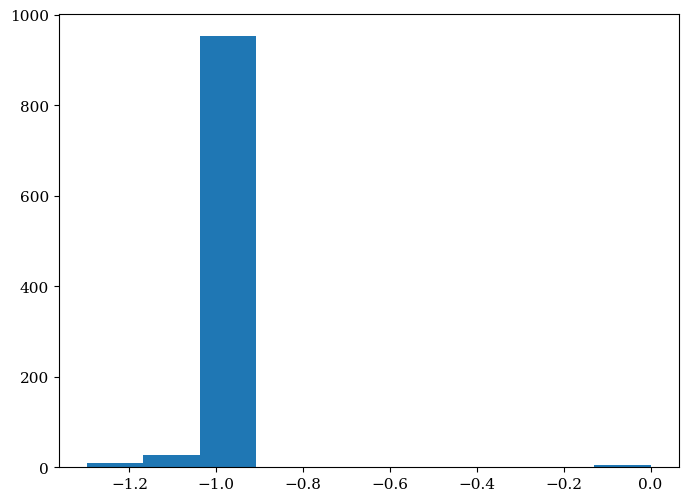}
\includegraphics[width=0.31\textwidth]{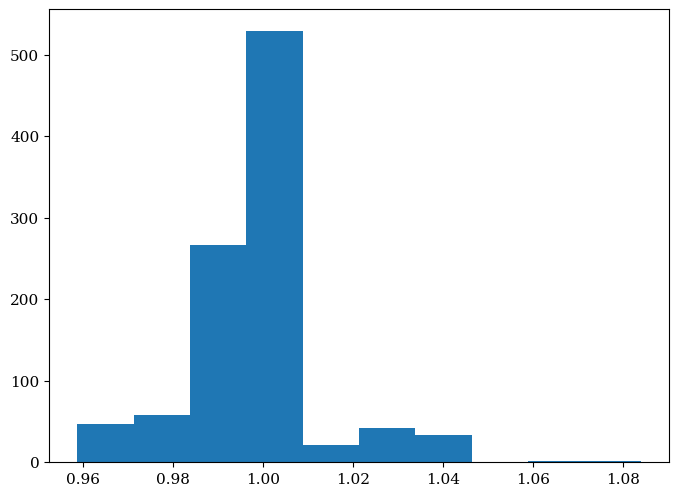}
\caption{Left: Distribution of the values of the Laplacian $\Delta d$
($\Delta_p d$ with $p=2$).
Middle: Distribution of the values of the p-Laplacian $\Delta_p d$, 
$p=8$.
Right: Distribution of the values of $|\nabla d|$. 
The computational domain is the interval $[-2,3]$. 
The values are centered around $-1$ for $\Delta_p d$ 
and $1$ for $|\nabla d|$.
}
\label{fig:1d_histogram}
\end{figure}

To verify the quality of the solutions, we also show in 
Fig. \ref{fig:1d_histogram} the distribution of the 
$p$-Laplacian $\Delta_p d(\x)$, $p=2,8$, for the solution to (\ref{pPoisson}), in the left and middle images, 
and the norm of the gradient $|\nabla_{\x} d(\x)|$, for the solution to (\ref{eik}), 
in the right image. 
Notice how the distributions are centered around $-1$ for the solution to the 
$p$-Poisson problem, and around $1$ for the solution to the eikonal equation. 

\subsubsection{2D example}
In 2D, we consider the function 
$$
f(x,y) = 1-x^2-y^2,
$$ 
whose zero level-set correspond to a unit circle. 
\\
The exact SDF to the unit circle is, of course, known and given by 
$$
1 - \sqrt{x^2+y^2}.
$$
\begin{figure}[!h!tbp!]
\centering
\includegraphics[width=0.45\textwidth]{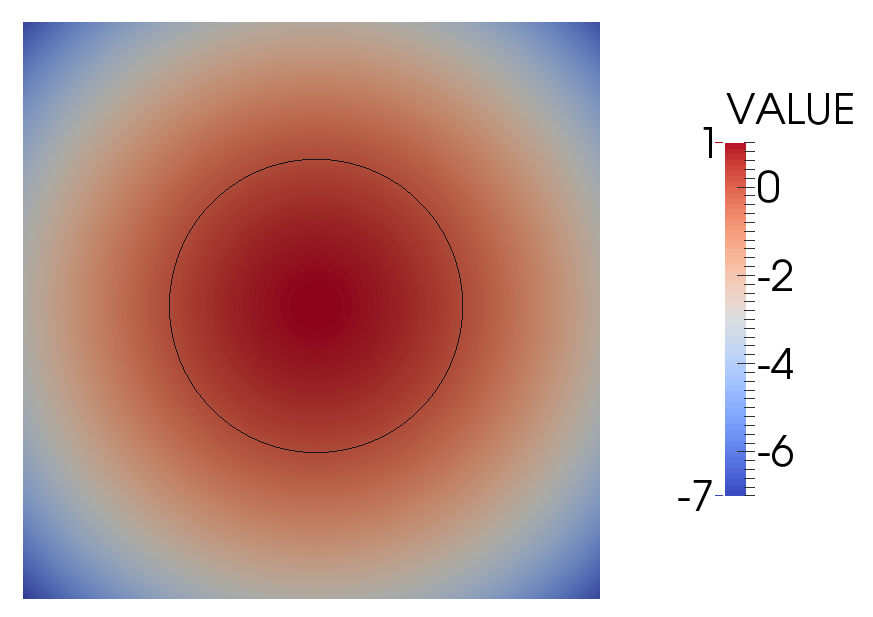}
\includegraphics[width=0.45\textwidth]{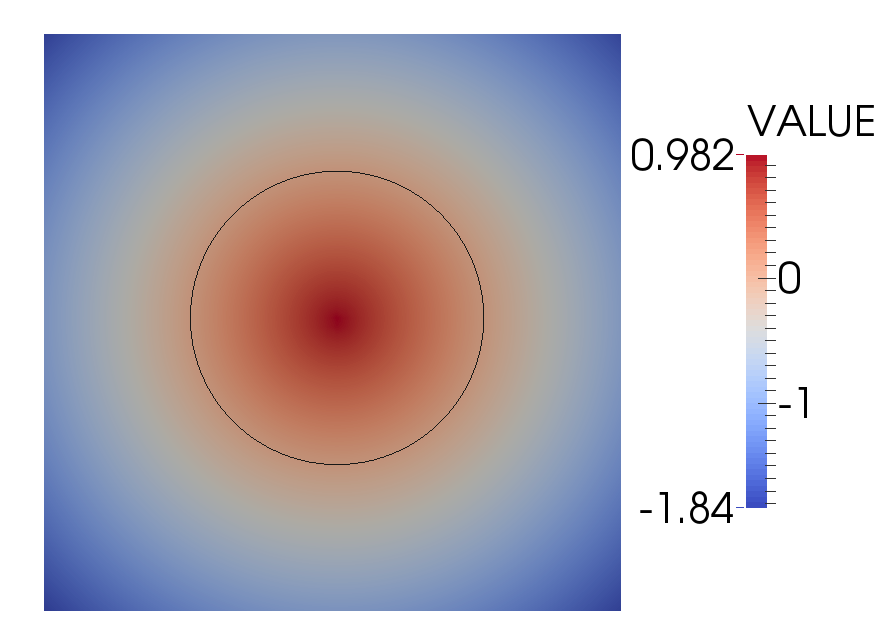}
\includegraphics[width=0.45\textwidth]{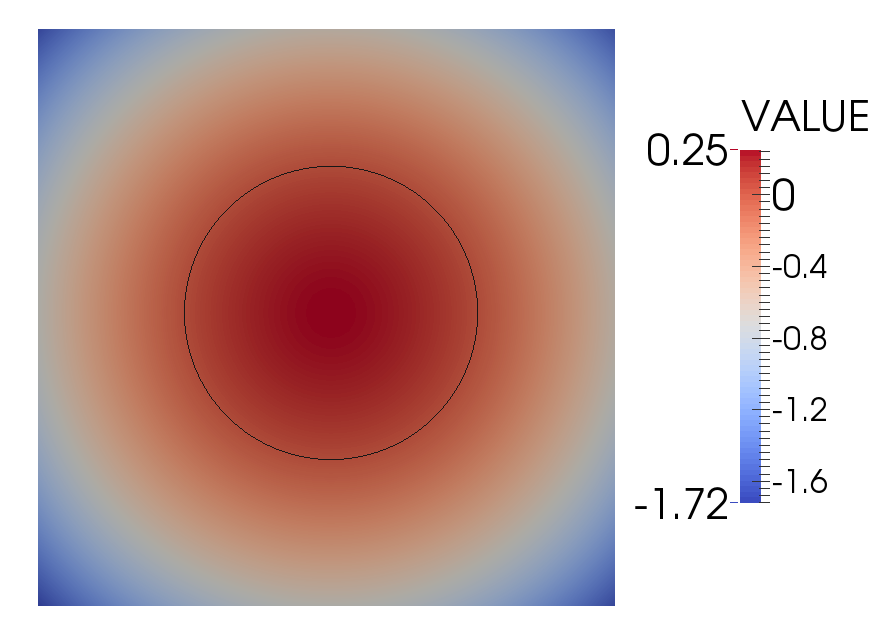}
\includegraphics[width=0.45\textwidth]{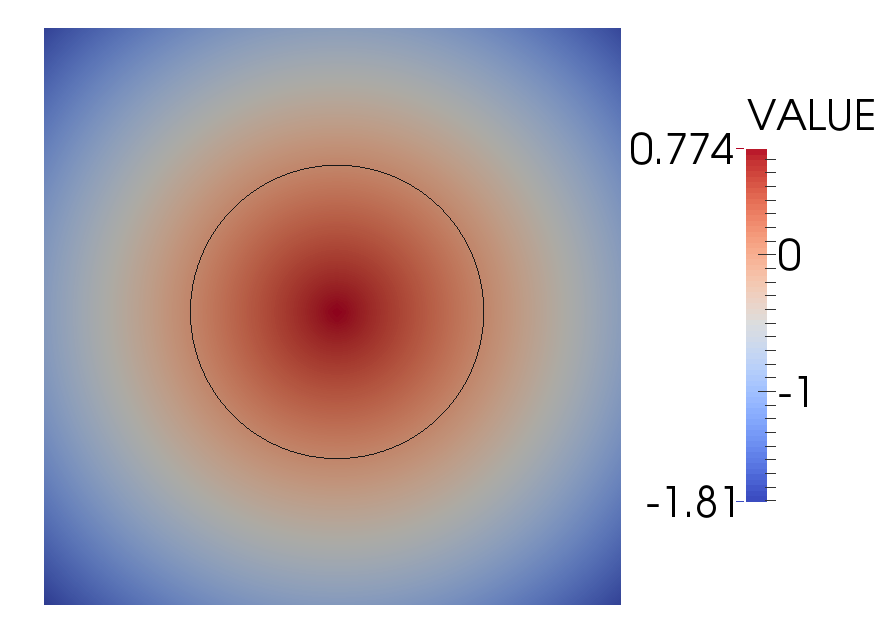}
\caption{Top row: 
Filled contour plot of $f(x,y)=1-x^2-y^2$ (left). 
Right image: SDF obtained by minimizing the loss (\ref{loss_eik}). 
Bottom row: 
Approximate SDF obtained by minimizing (\ref{loss_pPoisson}) 
for $p=2$ (left) and $p=8$ (right). 
Note how the zero level-set of $f$ is exactly preserved in all cases.}
\label{fig:2d_plot}
\end{figure}

Figure \ref{fig:2d_plot} illustrate the results obtained by our approach. 
The top row shows filled contour plots for the input function $f(x,y)$ and 
the solution to the eikonal equation (\ref{eik}) obtained by minimizing (\ref{loss_eik}). 
The bottom row show the solution to the $p$-Poisson problem for $p=2$ (left) and 
$p=8$ (right), respectively. 
In all cases, we used $15000$ iterations of ADAM. 
Notice again how the zero level-set of $f$ is exactly preserved. 

\subsubsection{3D example}
\begin{figure}[!h!tbp!]
\centering
\includegraphics[width=0.45\textwidth]{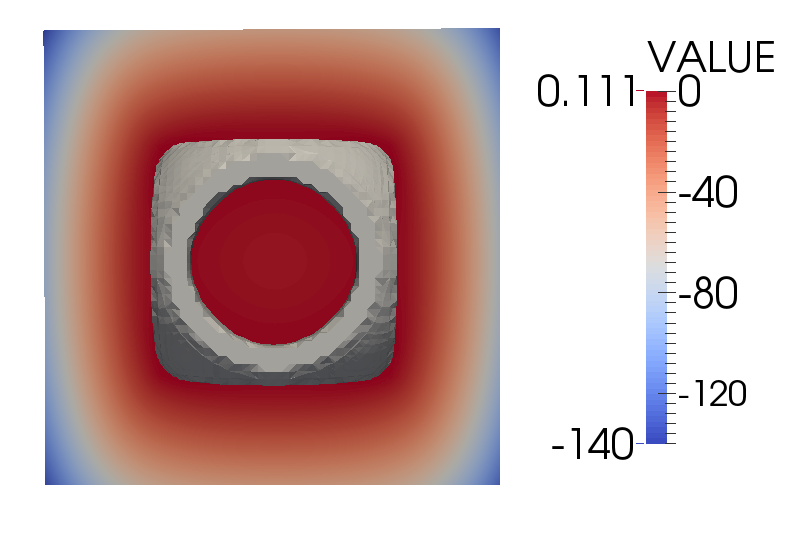}
\includegraphics[width=0.45\textwidth]{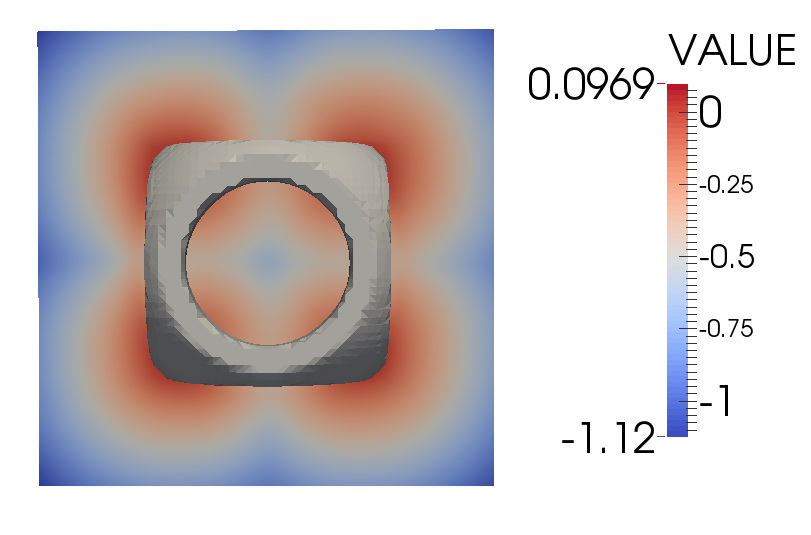}
\caption{Zero level-set and filled contour plot on a slice for 
the original implicit (left) and the SDF obtained by minimizing 
the loss (\ref{loss_eik}).}
\label{fig:3d_plot}
\end{figure}

We use a more complex model to illustrate our approach in 3D. 
See the left image in Fig. \ref{fig:3d_plot}. 
This object is built from "algebraic" primitives (a sphere, a box and three cylinders), 
and CSG operations implemented by R-functions \cite{pasko1995function}. 
The term "algebraic" refers here to the fact that we are not using 
distance-based primitives. 
For example, we use 
$$
1 - x^2 - y^2 - z^2, 
$$
and 
$$
1 - x^2 - y^2
$$
for a sphere and a cylinder, respectively. 
\\
Figure \ref{fig:3d_plot}, left image, shows with a filled contour plot 
on a given slice of the object that the function $f$ defining the object 
does not correspond to the distance to the surface $S$. 
\\
Figure \ref{fig:3d_plot}, right image, shows the solution to (\ref{eik}). 
It corresponds to the deep neural network obtained from minimizing 
the loss function (\ref{loss_eik}). 
We used $15000$ iterations of ADAM (similar to the 1D and 2D examples). 
Notice how the solution delivers a good approximation of the signed distance 
to the surface. Since the original implicit $f$ is embedded in the last layer of 
the deep neural network, the zero level-set of $f$ is exactly preserved. 

\section{Conclusion}
We proposed in this short note a technique for re-distancing an implicit surface. 
Given a function $f$ whose zero level-set defines a geometric surface $S$, we want to 
compute the signed distance to $S$ (or an approximation), while preserving exactly 
the zero level-set of $f$. 
We do that by embedding the function $f$ in the last layer of a deep neural network, 
which is then trained on a loss function derived from an eikonal equation. 
Alternatively, we propose a different loss function corresponding to the $p$-Poisson 
problem $\Delta_p u = -1$, whose solution delivers an approximation to the distance 
function. 

\bibliographystyle{plain}
\bibliography{sdf}

\end{document}